\documentclass{elsart}

\usepackage{graphicx}
\usepackage{amssymb}

\begin{document}

\begin{frontmatter}

\title{Anticipated synchronization in coupled chaotic maps with delays}

\author[M]{Cristina Masoller},
\ead{cris@fisica.edu.uy}
\author[Z]{Dami\'an H. Zanette}
\ead{zanette@cab.cnea.gov.ar}

\address[M]{Instituto de F\'{\i}sica, Facultad de Ciencias,
Universidad de la Rep\'ublica, Igu\'a  4225, Montevideo 11400,
Uruguay}

\address[Z]{Consejo Nacional de Investigaciones Cient\'{\i}ficas y
T\'ecnicas, Centro At\'omico Bariloche and Instituto Balseiro,
8400 Bariloche, R\'{\i}o Negro, Argentina}

\begin{abstract}
We study the synchronization of two chaotic maps with
unidirectional (master-slave) coupling. Both maps have an
intrinsic delay $n_1$, and coupling acts with a delay $n_2$.
Depending on the sign of the difference $n_1-n_2$, the slave map
can synchronize to a future or a past state of the master system.
The stability properties of the synchronized state are studied
analytically, and we find that they are independent of the
coupling delay $n_2$. These results are compared with numerical
simulations of a delayed map that arises from discretization of
the Ikeda delay-differential equation. We show that the critical
value of the coupling strength above which synchronization is
stable becomes independent of the delay $n_1$ for large delays.
\end{abstract}

\end{frontmatter}

\section{Introduction}

Time-delay systems have attracted a lot of attention in recent
years, in part due to the fact that multistability, i.e. the
coexistence of multiple attractors, is a common occurrence when
the delays are large --typically, much larger than the response
time of the system \cite{1}. Interest in multistability arises
because multistable systems play a key role in pattern recognition
processes \cite{2} and memory storage devices. By choosing
appropriate initial conditions, prescribed periodic solutions can
be stored as oscillatory patterns of a time-delay system
\cite{3,4,5}. Synchronization of chaotic time-delay systems has also
received attention, since it has potential applications to secure
communications \cite{6,7}. Perez and Cerdeira \cite{8} have shown
that, in low-dimensional chaotic systems, a hidden message can be
unmasked by the dynamical reconstruction of the chaotic signal
using nonlinear dynamical methods. Encrypting a message in the
chaotic output of a time-delay system has the advantage that the
dynamics is in this case high-dimensional (the dimension increases
linearly with the delay \cite{9}) but, in spite of this fact,
synchronization can be achieved by transmitting a single scalar
signal. Unfortunately, this method is not as secure as initially
expected, since it has been shown that by using a special
embedding space, the delay time can be identified, and the
message can be successfully unmasked \cite{10}. Coupled oscillators
with time delays in the coupling, which represent interactions
being transmitted at finite speed, have been extensively studied
as well \cite{11,12,13}.

Recently, a new effect of delayed coupling was reported by Voss
\cite{14,15}, who showed the existence of an anticipating
synchronization regime. In this regime, the slave system becomes
synchronized to the chaotic future state of the master system.
Anticipation occurs when the coupling is delayed, and results
from the interplay of memory effects and relaxation mechanisms.
Numerically, this regime was found in coupled semiconductor
lasers with optical feedback \cite{16}.

In this paper we investigate the existence and stability of
anticipated synchronization in coupled time-delay maps, which have
the advantage of allowing for analytical calculations. In the next
section, we introduce a system of two time-delay coupled maps in
a master-slave configuration, and present analytical results on
the stability of anticipated and retarded synchronization for
generic maps. In Section 3, we apply the results to a delay map
that arises from the discretization of the Ikeda
delay-differential equation. Section 4 presents a summary and the
conclusions.

\section{Master-slave coupled delay maps}

We consider a one-dimensional map of the form
\begin{equation} \label{x}
x_{n+1}= b x_n + f(x_{n-n_1}).
\end{equation}
For $|b|<1$, the first term in the right-hand side represents a
relaxation mechanism. Under its sole action $x_n$ would
asymptotically vanish. This relaxation, however, competes with
the effect of the nonlinear function $f(x)$, which has the form
of a time-delayed feedback with delay $n_1$.

The (master) map (\ref{x}) is used to partially drive the
evolution of a new (slave) system $y_n$. The dynamics of this
system is, in principle, the same as for $x_n$, except that a part
of the nonlinear component is replaced by the evolution of $x_n$,
namely
\begin{equation} \label{y}
y_{n+1}= b y_n + (1-\eta) f(y_{n-n_1})+ \eta f(x_{n-n_2}).
\end{equation}
The parameter $\eta \in [0,1]$ measures the strength of the
driving action. Note that $x_n$ enters the dynamics of $y_n$ with
a delay $n_2$. This form of unidirectional time-delay coupling is
not only an extension to maps of the configuration proposed by
Voss for continuous-time systems \cite{10}, but also generalizes
the master-slave interaction by introducing a parameter ($\eta$)
that controls its strength. Voss's configuration corresponds to
the extreme value $\eta=1$.

The structure of Eqs. (\ref{x}) and (\ref{y}) suggests that, if
driving is effective enough, $y_n$ may become synchronized to the
driving signal, in the form
\begin{equation} \label{sync}
y_n = x_{n-n_2+n_1}.
\end{equation}
If this synchronization is realized, the dynamics of $y_n$
coincides with that of $x_n$ up to a time shift $n_1-n_2$. If
$n_1>n_2$ the slave system anticipates the evolution of the
master system, and we have anticipated synchronization. On the
other hand, if $n_1 < n_2$ the synchronization is retarded.

Linear stability of the synchronized state (\ref{sync}) can be
analyzed in the usual way. Taking $y_n = x_{n-n_2+n_1}+\delta
y_n$, the evolution of the deviation $\delta y_n$ in the limit of
infinitesimal deviations is readily derived from (\ref{y}):
\begin{equation} \label{delta}
\delta y_{n+1}=b \delta y_n+ (1-\eta) f'(x_{n-n_2}) \delta
y_{n-n_1},
\end{equation}
with $f'\equiv df/dx$. Note that, if $|b|<1$ and for $\eta=1$,
the deviation decreases exponentially with time  and
synchronization is therefore stable. In view of well-known
properties of synchronization in coupled maps without delays, it
is possible to advance that, if the dynamics of $x_n$ is chaotic,
the synchronized state (\ref{sync}) will be stable above a
certain threshold $\eta_c<1$. On the other hand, if $x_n$ is
nonchaotic, synchronization should be stable for any $\eta>0$.

For $n_1\neq 0$, Eq. (\ref{delta}) can be formally integrated by
introducing a linear $(n_1+1)$-dimensional map for a variable
${\bf z}_n= (z^0_n, z^1_n,\dots,z^{n_1}_n)$, with $z^k_n= \delta
y_{n-k}$. This equivalent map is given by
\begin{equation}
{\bf z}_{n+1}= M_n {\bf z}_n,
\end{equation}
where the elements of the matrix $M_n$ read
\begin{equation} \label{mat}
M_n^{i,j} = \left\{
\begin{array}{ll}
b & \mbox{if $i=j=0$,} \\
(1-\eta) f'(x_{n-n_2}) & \mbox{if $i=0, j=n_1$,} \\
1 & \mbox{if $i=j+1$ $(j=0,\dots,n_1-1)$,} \\
0 & \mbox{otherwise.}
\end{array}
\right.
\end{equation}
The deviations given by Eq. (\ref{delta}) can then be obtained
from the solution of the equivalent map,
\begin{equation} \label{U}
{\bf z}_n = U_n {\bf z}_0= M_{n-1}M_{n-2} \cdots M_1M_0
{\bf z}_0.
\end{equation}
Thus, synchronization is linearly stable if all the eigenvalues
of the evolution matrix $U_n$ vanish for $n \to \infty$. For
$\eta=1$ the eigenvalues are $(b^n,0,\dots,0)$. For $\eta<1$, on
the other hand, $U_n$ and its eigenvalues must be calculated
numerically, since they involve the successive states of the
variable $x_n$, cf. Eq. (\ref{mat}). It is interesting to point
out that stability is determined by the asymptotic properties of
$U_n$ in the limit $n\to \infty$, where the driving system $x_n$
has exhaustively explored the accessible region of phase space.
In this limit, thus, the delay $n_2$ of Eq. (\ref{y}) plays no
role in the determination the eigenvalues of $U_n$. In
consequence, $n_2$ is irrelevant to the stability of the
synchronized state.

\section{Application to the Ikeda delay map}

In this section we present numerical results corresponding to maps
(\ref{x}) and (\ref{y}), with $f(x)= a \sin x$. With this choice
of $f(x)$, we refer to Eq. (\ref{x}) as the Ikeda delay map, since
it arises from the discretization of the Ikeda delay-differential
equation \cite{1,17},
\begin{equation} \label{Ik}
 \dot x=-x+\alpha \sin x (t-\tau).
\end{equation}
This equation was introduced to describe the dynamics of an
optical bistable resonator, where the finiteness of the speed of
light is relevant, and therefore require a description that
incorporates explicitly the round-trip time of light in an
optical cavity. Physically, $x$ is the phase lag of the electric
field across the resonator, $\alpha$ is the laser intensity
injected into the system, and $\tau$ is the round trip time of
the light in the resonator.

Writing $\dot x \equiv [x(t+\Delta t)-x(t)]/ \Delta
t$, $t\equiv n\Delta t$, and $\tau \equiv n_1 \Delta t$, the
Ikeda delay map is obtained from Eq. (\ref{Ik}) with $b\equiv 1-
\Delta t$, $a\equiv  \alpha \Delta t$, and $x_n \equiv
x(t)$. The Ikeda delay map is not to be confused with the
discrete Ikeda map \cite{1,18}, which is obtained from (\ref{Ik}) by
discretizing time in units of $\tau$ in the singular limit where
the delay-to-response time ratio diverges.

\begin{figure}
\centering
\resizebox{\columnwidth}{!}{\includegraphics{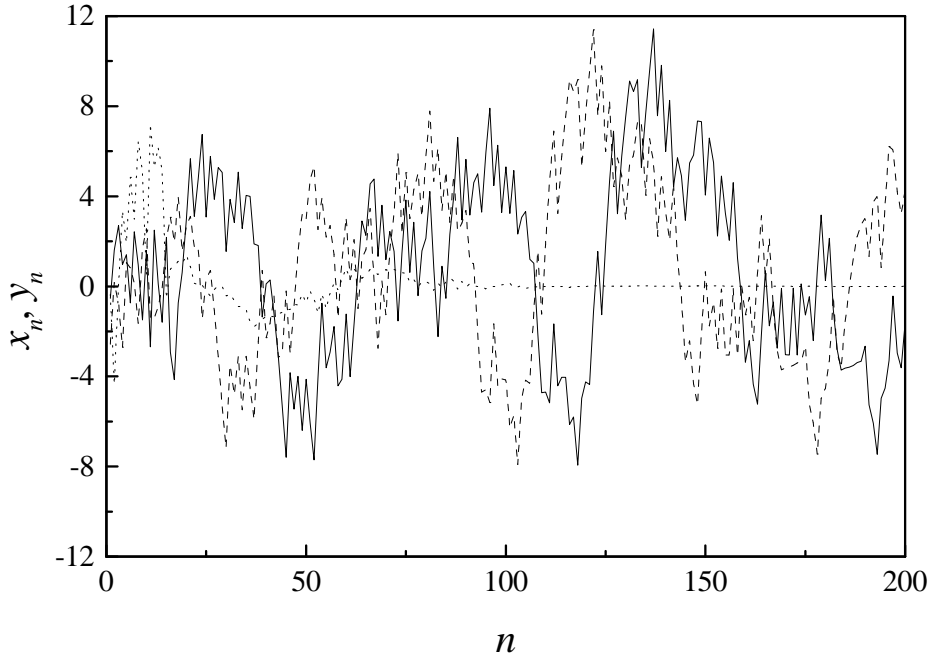}}
\caption{Time series of the maps (\ref{x}) ($x_n$, solid line) and
(\ref{y}) ($y_n$, dashed line) with $f(x)=a \sin x$, for $a=3$,
$b=0.9$, $\eta=0.833$, $n_1=15$, and $n_2=0$. The dotted line
stands for the difference $x_{n+n_1-n_2}- y_{n}$.}
\label{f1}
\end{figure}

In the following, we focus the attention on the case $b=0.9$. Our
numerical calculations are restricted to random initial condition
in $(-\pi,\pi)$ both for the master and the slave system. In Fig.
\ref{f1} we illustrate anticipated synchronization, for $n_1=15$
and $n_2=0$. The slave system (dashed line) anticipates in $15$
steps the state of the master system (solid line). After a
transient, the difference $x_{n+n_1-n_2}-y_{n}$ (dotted line)
decays to zero. We have verified that this behavior is
independent of the value of $n_2$.

The critical value $\eta_c$ of the coupling strength, above which
the synchronized state is stable, has been calculated from the
analytical results of the previous section by means of a numerical
evaluation of the matrix $U_n$ in Eq. (\ref{U}). Since the delay
$n_2$ is irrelevant to this calculation, we can take $n_2=0$ in
Eq. (\ref{mat}). From a given initial condition for the master
system, the matrix $U_n$ is evaluated for successive values of
$n$. In order to avoid the calculation of its eigenvalues, which
is specially troublesome for large $n_1$, $U_n$ is multiplied at
each step by a randomly generated vector, ${\bf u}_n$, of unitary
modulus. If, for a given value of $\eta$, the modulus of the
product remains above a certain upper threshold $H$ for
sufficiently large $n$, $|U_n {\bf u}_n|>H$ for $n>n_{\max}$,
synchronization is considered unstable and $\eta_c>\eta$. If, on
the other hand, $|U_n {\bf u}_n|<h$ for $n>n_{\max}$, where $h$
is a suitable lower threshold, synchronization is considered
stable and $\eta_c <\eta$. With this criterion, $\eta_c$ can be
found by decimation within the interval $(0,1)$ up to a certain
previously fixed precision. In our calculations, we have taken
$H=10^{10}$, $h=10^{-10}$, and $n_{\max} \sim 10^4$.

\begin{figure}
\centering
\resizebox{\columnwidth}{!}{\includegraphics{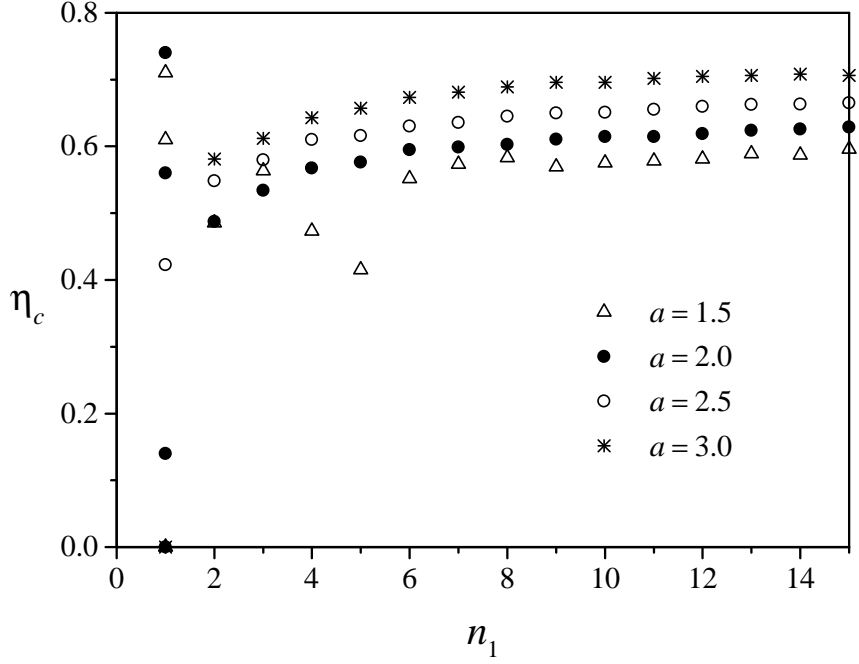}}
\caption{Critical
value of the coupling strength above which the synchronized state
is stable, calculated by means of a numerical evaluation of the
matrix $U_n$ in Eq. (\ref{U}), as explained in the text.}
\label{f2}
\end{figure}

Figure \ref{f2} shows $\eta_c$ as a function of the delay $n_1$
and for several values of the parameter $a$. Note that for small
$n_1$ --specifically, for $n_1=1$-- the critical value $\eta_c$
can vanish, which indicates that the master evolution is
nonchaotic. Notice, furthermore, that for some values of $a$ and
$n_1=1$ we have plotted more than one value of $\eta_c$. This is
due to the fact that, in such cases, the master system is
multistable --with, typically, two chaotic attractors. The value
of $\eta_c$ depends thus on the initial conditions for both $x_n$
and $y_n$. Whereas the behavior is quite irregular for small
$n_1$, for larger delays the critical coupling becomes
practically independent of $n_1$. Though it has not been possible
to prove this analytically, we conjecture that $\eta_c$
approaches a finite limit below unity as $n_1 \to \infty$. The
dependence on $a$ in the investigated interval is also moderately
weak. As expected, since the dynamics of the Ikeda delay map
becomes more irregular as $a$ grows --i.e. the Lyapunov exponent
is higher-- $\eta_c$ increases accordingly.

\begin{figure}
\centering
\resizebox{\columnwidth}{!}{\includegraphics{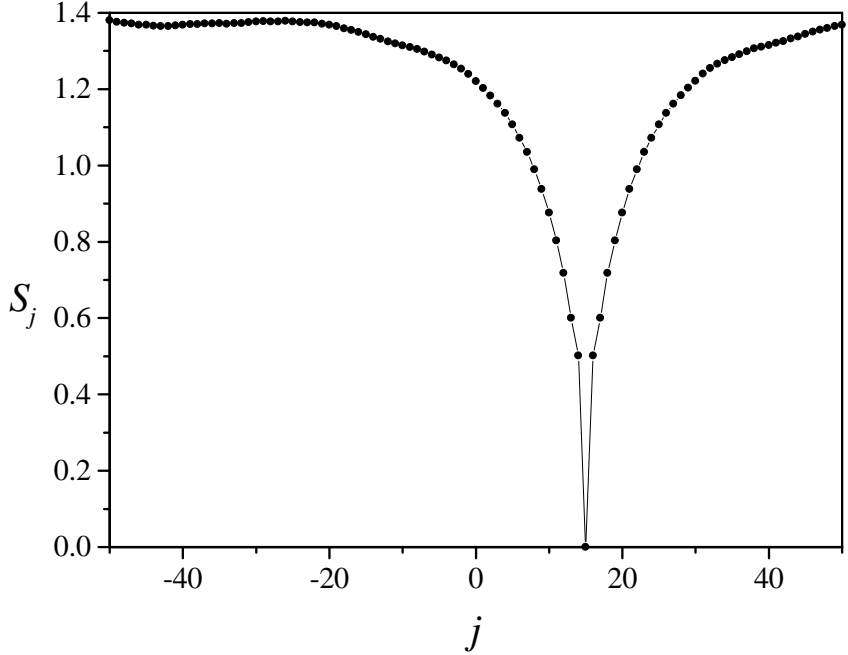}}
\caption{ Similarity function, Eq. (\ref{S}), for the
same parameters as in Fig. \ref{f1} ($n_1-n_2=15$).}
\label{f3}
\end{figure}

The degree of anticipated of retarded synchronization can be
quantified by calculating the similarity function $S_j$, defined
as \cite{19}:
\begin{equation} \label{S}
S_j^2={{\langle [ x_{n+j}-y_{n}]^2\rangle } \over{[ \langle
x_n^2\rangle \langle y_n^2\rangle ]^{1/2}}} .
\end {equation}
If $x_n$ and $y_n$ are independent time series with similar mean
value and dispersion, the average of their square difference  is
of order $2 \langle x_n^2\rangle \approx 2 \langle y_n^2\rangle$,
and thus $S_j \approx \sqrt{2} \approx 1.4$. If, on the other
hand, there is perfect anticipated or retarded synchronization,
the difference $x_{n+n_1-n_2}- y_{n}$ vanishes, and $S_{n_1-n_2}
=0$. In Fig. \ref{f3}, the similarity function is shown  for the
same parameters as in Fig. \ref{f1}.

\begin{figure}
\centering
\resizebox{\columnwidth}{!}{\includegraphics{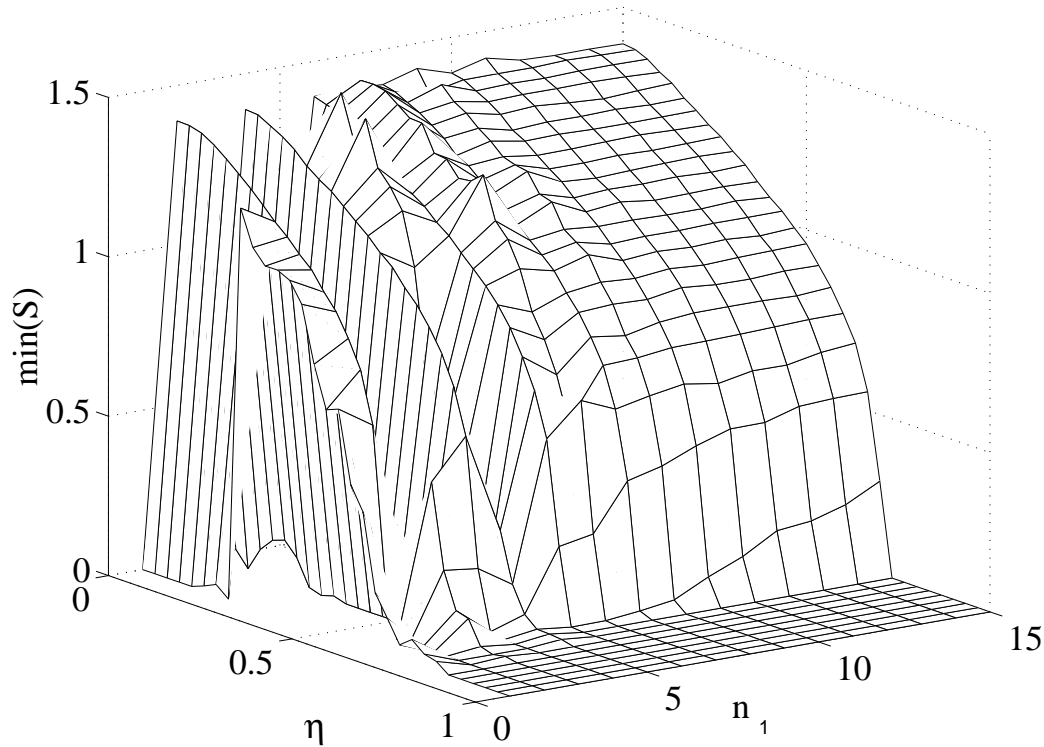}}
\caption{ Minimum of the similarity function in the parameter
space ($\eta,n_1$), with all the other parameters as in Fig.
\ref{f1}. The minimum was calculated by averaging over $50$
trajectories with different initial conditions.}
\label{f4}
\end{figure}

Figure \ref{f4} shows  $\min (S)$ --i.e. the minimum value of
$S_{n_1-n_2}$ recorded during sufficiently long realizations of
the evolution for a given set of parameters-- in the
$(n_1,\eta)$-plane, with all the other parameters as in Fig.
\ref{f1}. The region where anticipated synchronization occurs,
$\min (S) \approx 0$, is clearly visible for large values of
$\eta$. The position of its boundary is in good qualitative
agreement with the values of $\eta_c$ shown in Fig. \ref{f2} for
$a=3$.

\section{Conclusion}

We have extended the study of anticipated synchronization,
advanced by Voss for unidirectionally coupled differential
equations with time delays \cite{14,15}, to delayed coupled
chaotic maps. While the nature of anticipated synchronization of
maps and differential equations is the same, delay discrete-time
dynamics admits an analytical treatment which cannot be carried
out for continuous-time systems. In fact, ordinary differential
equations with finite time delays constitute an
infinite-dimensional problem. On the other hand, since times
delays in maps must be discrete, the dimensionality of the
problem remains finite.

Taking advantage of this situation, we have analytically studied
the stability of anticipated and retarded synchronization in a
generic master-slave configuration. In the absence of coupling,
master and slave dynamics are identical and involve an intrinsic
delay $n_1$. Coupling consists in the replacement of a part of the
slave dynamics by that of the master system, with a delay $n_2$.
We have shown that the stability of synchronization is
independent of $n_2$. The structure of the linearized problem,
Eqs. (\ref{delta}-\ref{U}), suggest meanwhile a strong --though
not transparent-- dependence on the Lyapunov exponent of the
master system, as expected. In practice, the linearized problem
has to be treated numerically, but it only involves the
realization of the master system and the successive application
of the linear-evolution operator, being thus a purely algebraic
process.

These results have been applied to the Ikeda delay map, which
derives from the application of the Euler integration scheme to
the Ikeda delay-differential equation. We have calculated the
critical coupling intensity above which synchronization is
stable, as a function of the delay $n_1$ and of a parameter $a$
that controls the chaotic dynamics of the map. It is found that,
whereas for small values of $n_1$ the critical coupling can vary
considerably --due to the irregular appearance and disappearance
of chaotic and nonchaotic attractors-- the dependence for large
$n_1$ is much smoother. In fact, as $n_1$ grows, the critical
coupling seems to approach a constant value. The dependence with
the dynamical parameter $a$ is also moderate in the considered
range.

In this work we have focused the attention on exact anticipated
synchronization. However, it has been previously shown that
approximate anticipated synchronization is possible in coupled
differential equations, even in the absence of intrinsic delays
\cite{15}. The study of approximate anticipated synchronization
in coupled maps constitutes therefore a line for future work. In
particular, it would be interesting to investigate in detail the
connection between the degree of synchronization, and the
irregularity of the dynamics --as measured by the Lyapunov
exponents. For delay differential equations, the number of
positive Lyapunov exponents and the fractal dimension increase
linearly with the delay, while the metric entropy remains roughly
constant \cite{9,20,21}. We therefore conjecture that the metric
entropy might be a good indicator of the possibility of
synchronizing with anticipation, and thus predicting, chaotic
dynamics.

\section*{Acknowledgements}

This work was supported by Proyecto de Desarrollo de Ciencias
B\'asicas (PE\-DE\-CI\-BA) and by Comisi\'on Sectorial de
Investigaci\'on Cient\'{\i}fica (CSIC), U\-ru\-guay.


\begin{thebibliography}{00}

\bibitem{1} K. Ikeda, K. Matsumoto, Physica D 29  (1987) 223-235.

\bibitem{2} S. Kim, S. H. Park,  C. S. Ryu, Phys. Rev. Lett.
79  (1997) 2911-2914.

\bibitem{3} B. Mensour, A. Longtin, Phys. Lett. A 205 (1995) 18-24

\bibitem{4} J. Foss, A. Longtin, B. Mensour,  J. Milton, Phys.
Rev. Lett.  76 (1996) 708-711.

\bibitem{5} B. Mensour,  A. Longtin, Phys. Rev. E 58
(1998) 410-422.

\bibitem{6} M. J. Bunner, W. Just, Phys. Rev. E 58 (1998) R4072-R4075

\bibitem{7} R. He, P. G. Vaidya, Phys. Rev. E 59 (1999) 4048-4051

\bibitem{8} G. Perez, H. A. Cerdeira, Phys. Rev. Lett. 74 (1995) 1970-1973.

\bibitem{9} J. D. Farmer, Physica D 4 (1982) 366-393.

\bibitem{10} C. Zhou, C.-H. Lai, Phys. Rev. E 60 (1999) 320-323

\bibitem{11} M. Y. Choi, H. J. Kim, D. Kim,  H. Hong, Phys. Rev.
E  61 (2000) 371-381.

\bibitem{12} D. Zanette, Phys. Rev. E 62 (2000) 3167-3172.

\bibitem{13} D. V. Raman Reddy, A. Sen, G. L. Johnston,
Phys. Rev. Lett.  85 (2000) 3381-3384.

\bibitem{14} H. U. Voss, Phys. Rev. E 61 (2000) 5115-5119.

\bibitem{15} H. U. Voss, Phys. Lett. A 279 (2001) 207-214.

\bibitem{16} C. Masoller, Phys. Rev. Lett. 86 (2001) 2782-2785.

\bibitem{17} K. Ikeda, H. Daido, and O. Akimoto,
Phys. Rev. Lett. 45 (1980) 709-712.

\bibitem{18} P. Mandel, Theoretical Problems in Cavity Nonlinear
Optics, Cambridge University Press, New York, 1997, and references
therein.

\bibitem{19} M. G. Rosenblum, A. S. Pikovsky, J. Kurths,
Phys. Rev. Lett. 78  (1997) 4193-4196.

\bibitem{20} M. Le Berre, E. Ressayre, A. Tallet,
H. M. Gibbs, Phys. Rev. Lett. 56 (1986) 274-277.

\bibitem{21} C. Masoller, Chaos 7 (1997) 455-462.

\end{thebibliography}
\end{document}